\documentclass[runningheads,a4paper]{llncs}

\usepackage{amssymb}
\usepackage{graphicx}
\usepackage{subcaption}
\usepackage{url}
\urldef{\mailsa}\path|{christos.rodosthenous, loizos.michael}@ouc.ac.cy|
\newcommand{\keywords}[1]{\par\addvspace\baselineskip
\noindent\keywordname\enspace\ignorespaces#1}

\begin{document}

\mainmatter 

\title{Web-STAR: Towards a Visual Web-Based IDE for a Story Comprehension System}

\titlerunning{Towards a Visual Web-Based IDE for a Story Comprehension System}

%
%
\author{Christos Rodosthenous \and Loizos Michael}
\authorrunning{Christos Rodosthenous \and Loizos Michael}

\institute{Open University of Cyprus\\
\mailsa\\
}

%
%

\toctitle{Lecture Notes in Computer Science}
\tocauthor{Authors' Instructions}
\maketitle

\begin{abstract}
In this work, we present Web-STAR, an online platform for story understanding built on top of the STAR (STory comprehension through ARgumentation) reasoning engine. This platform includes a web-based IDE, integration with the STAR system and a web service infrastructure to support integration with other systems that rely on story understanding functionality to complete their tasks. The platform also delivers a number of ``social" features like public story sharing with a built-in commenting system, a public repository for sharing stories with the community and collaboration tools that can be used from both project team members for development and educators for teaching. Moreover, we discuss the ongoing work on adding new features and functionality to this platform.
\end{abstract}

\section{Introduction}
\label{section:introduction}
It is important for a logic-based system to have an intuitive interface that allows both expert and non-expert users to use it for preparing programs, debugging them and learning to use the provided features. In this work, we focus on such a system, the STAR (STory comprehension through ARgumentation) reasoning engine \cite{Diakidoy2015,Diakidoy2014} that deals with the problem of story comprehension and we present a platform built on top of it.

Understanding stories, is one of the many tasks humans can perform well. Machines are still not capable of performing such a task as well as humans do, since it requires the ability to infer about states and events that are not explicitly described in the text \cite{Mueller:2003:SUT:1119239.1119246}. This process is called commonsense reasoning and it requires knowledge about the world we live in (background knowledge).

Classical logic and variance has been used to represent commonsense knowledge and to give semantics for the reasoning process. Recent work \cite{Kakas.etal_2016_ArgumentAndCognition} has suggested the use of argumentation \cite{Bench-Capon2007,besnard2008elements} as a more appropriate framework for the representation and reasoning with commonsense knowledge.

In this work, we focus on the design and development of the \textbf{Web-STAR} platform. This platform includes a web-based integrated development environment (IDE) that presents a personalized environment for each user with tools for writing, executing, debugging stories using the STAR system and visualizing their output. Moreover, the IDE is the basis of a community, where people can share stories, comment on them and reuse other community created stories. Under the same umbrella, a web service is also available for integrating other systems with the Web-STAR platform.

For increasing the usage and friendliness of the Web-STAR IDE, a number of design options are considered. These design options allow novice users to prepare their stories and add background knowledge in a more graphical or visual way, e.g., by using a directed graph representation of background knowledge and by using wizards to add questions and story related information.

In the following sections, we present the current state in story understanding systems and web IDEs followed by a presentation of the STAR system that is the underlying engine of the Web-STAR platform. Next, the Web-STAR platform is presented using a realistic scenario of using its IDE. In the final section, new features and additions to the Web-STAR platform are discussed as part of the ongoing work on the platform.

\section{Related Work}
\label{section:related_work}
In this section, we present related work on both story understanding systems and web-based IDEs, accompanied with examples of both general language and logic-based IDEs.

\subsection{Story Understanding Systems}
\label{subsection:story_understanding_systems}
Currently, several systems have been developed to deal with the problem of story comprehension. Starting from the 70's, Charniak presented two systems; a story comprehension model for answering questions about children stories, by relating stories to real-world background knowledge \cite{Charniak1972} and a system for answering questions about simple stories dealing with painting \cite{Charniak1973}. There is also work on Deep Read \cite{Hirschman1999}, an automated reading comprehension system that accepts stories and answers questions about them. More recent attempts, include work of Mueller \cite{mueller2007modelling}, on a system for modeling space and time in narratives about restaurants and Genesis \cite{winston:OASIcs:2015:5290}, a system for modeling and exploring aspects of story understanding using stories drawn from sources ranging from fairy tales to Shakespeare's plays. 

In terms of technical skills and expertise needed to use these systems, the majority of them use a command line interface (CLI), except Genesis that already has a visual interface. The lack of a visual environment makes it difficult for non-experts to setup and use these systems without prior programming knowledge and explicit knowledge of the specific system. Furthermore, the majority of these systems rely on external tools (text editors) for editing the source code and lack basic functionality that an IDE can easily provide (i.e., code folding, syntax highlighting, etc.).

\subsection{Web-Based IDEs}
\label{subsection:web_ides}
Developments in web technologies gave birth to numerous web-based IDEs, some of which are now considered as mainstream IDEs for developing applications. These are systems that are available through a web browser, they do not rely on specific hardware or software stack and are Operating System agnostic, e.g., Cloud9 \cite{cloud9website}, Codiad \cite{codiadwebsite}, ICEcoder \cite{icecoderwebsite}, codeanywhere \cite{codeanywherewebpage} and Eclipse Che\cite{eclipsechewebsite}.

These web-based IDEs allow users to write code in an online source code editor using their programming language of choice. They also provide code folding, code highlighting and auto-complete functionality, built-in their source code editors. Moreover, some of them provide online code execution functionality with access to an underlying virtual machine and thus access to the shell.

There are also IDEs concentrated on logic programming, like SWISH (SWI-Prolog for Sharing) \cite{Wielemaker2015} and IDP Web-IDE \cite{Dasseville2015}. SWISH is a web front-end for SWI-Prolog which is used to run small Prolog programs for demonstration, experimentation and education. IDP Web-IDE is an online front-end for IDP, a Knowledge Base System for the FO($\cdot$) language.

\section{The STAR System}
\label{section:star_system}
Before presenting the Web-STAR IDE, it is important to have an overview of the STAR system, the underlying engine used for story comprehension, and its semantics. The STAR system, is based on the well established argumentation theory in Artificial Intelligence \cite{Bench-Capon2007,Baroni:2011:RIA:2139707.2139708}, uniformly applied to reason about actions and change in the presence of default background knowledge \cite{Diakidoy2015}. The STAR system follows guidelines from the psychology of comprehension, both for its representation language and for its computational mechanisms for building and revising a comprehension model as the story unfolds.

The system takes as input a story written in symbolic predicate form using time-stamped facts, background knowledge in the form of association rules (e.g., \texttt{bird(X) implies fly(X)}) accompanied by priorities between them, and questions on the story facts. Each question can be asked as part of several distinct sessions, to check if there has been a revision in the comprehension model of the system as the story progresses. The output of the system includes the comprehension model and answers to the multiple choice questions posed for each session.

The STAR system is a Prolog application that runs in any environment which supports SWI-prolog \cite{wielemaker_schrijvers_triska_lager_2012}. It requires users to setup the Prolog environment, call the application's core engine file and select a domain file to load. Users can also choose what information is reported by the system during its execution.

A domain file follows the STAR system syntax and comprises two main parts; The ``story and questions part" that includes a series of sessions for representing the story narrative up to a scene (session), a set of questions up to that point, and the ``background knowledge part" that includes knowledge rules about the context of the story.

In the following paragraphs, we present the semantics of the STAR system using the following story in natural language:\\
\begin{quote}
Ann rang the doorbell. Mary, who was in the flat watching TV, got up from her chair and walked to the door. She was afraid. Mary looked through the keyhole. Mary saw Ann, her flatmate, at the door. Mary opened the door and asked Ann why she did not use her keys to come in the flat? Ann replied that she is upset that she did not agree to come with her at the shops. She wanted to get her up from her chair in front of the TV.
\end{quote}

The first part of the file encodes the various observations captured in the story with time-points for each one of them. The story content of a session is given by a set of observation statements of the form \texttt{s(\#N) $::$ \#GroundLiteral at \#Time-Point}, where \texttt{\#N} is a non-negative integer representing the session. A literal \texttt{\#Literal} is either a concept \texttt{\#Concept} or its negation -\texttt{\#Concept} (i.e., the symbol for negation is ``-"), where a concept \texttt{\#Concept} is a predicate name along with associated variables or constants for the predicate's arguments.

For example, the above short story could be represented as:

\begin{quote}
\texttt{s(0) $::$ person(ann) at always.}\\
\texttt{s(0) $::$ person(mary) at always.}\\
\texttt{s(0) $::$ request(go\_to(shops)) at always.}\\
\texttt{s(0) $::$ request(donate(money)) at always.}\\
\texttt{s(0) $::$ ring(ann,doorbell) at 2.}\\
\texttt{s(0) $::$ in\_flat(mary) at 3.}\\
\texttt{s(0) $::$ watch(mary,tv) at 3.}\\
$\dots$\\
\end{quote}

The story plot is presented in a single session, thus the \texttt{s(0)} notation in front of every observation statement. The first observations state that``ann" and``mary" are instances of type``person" and that holds at every time-point. After the observation statements, the questions are posed. A question in the STAR system follows this notation: \texttt{q(XX) $??$ PREDICATE1 at TIME-POINT; PREDICATE2 at TIME-POINT.}, where XX is a non-negative integer representing the number of the question and the ``;" separates the possible answers to that question. For the above story two example questions could be:

\begin{quote}
\% Question 1: Does Ann have the door keys?\\
\texttt{q(1) $??$ has(ann,doorkeys) at 1.} \% True or False type question\\

\% Question 2: Why did Mary walk to the door?\\ \% Possible answers: Because she wants to find out who is at the door or because she wants to open the door\\

\texttt{q(2) $??$ wants(mary,find\_out\_who\_at(door)) at 4;\\
\-\hspace{1.25cm} wants(mary,open(door)) at 4.}

\end{quote}

The second part of the domain file, includes the background knowledge required to comprehend the story. Background knowledge is represented in terms of associations between concepts. The STAR system supports three types of rules, property, causal, preclusion, of the following form respectively:
\begin{quote}
\texttt{p(\#N) :: \#Body implies \#Literal.}\\
\texttt{c(\#N) :: \#Body causes \#Literal.}\\
\texttt{r(\#N) :: \#Body precludes \#Literal.}
\end{quote}
The body of the rule is in the form of \texttt{\#Body = true | \#Literal | \#Literal, \#Body}. Background knowledge can also contain priority (i.e., relative strength) statements between the rules. Priorities are encoded in the form \#P1 $>>$ \#P2., where \#P1 and \#P2 are the names of two rules. The following is an example of such background knowledge used in the above story:

\begin{quote}
\% Walking to the door implies wanting to open the door, unless one is afraid.\\
\texttt{p(22) $::$ walk\_to(Person,door), in\_flat(Person) implies \\\-\hspace{1.29cm} wants(Person,open(door)).}\\
\texttt{p(23) $::$ afraid(Person), in\_flat(Person) implies \\
\-\hspace{1.29cm} -wants(Person,open(door)).}\\
\texttt{p(23) $>>$ p(22).} \% this captures the precedence of p(23) over p(22)\\
\% Seeing a flatmate at the door causes one to want to open the door.
\texttt{c(33) $::$ see\_at(Person,Other,door), flatmate(Person,Other) causes\\\-\hspace{1.30cm}  wants(Person,open(door)).}\\
\texttt{c(33) $>>$ p(23).} \% even if one is afraid.\\
$\dots$
\end{quote}

Before completing the story writing process, users need to define which questions will be answered in each session and which domain concepts should be visible on the screen as the system constructs its comprehension model. This is done using a session statement at the beginning of the story file:
\begin{quote}
\% session(s(\#N),\#Questions,\#Visible).\\
\texttt{session(s(0),[q(1),q(2)],all).}
\end{quote}

Users also need to declare the concepts that are fluents, i.e., persist over time and change across time. This is done using a statement of the form: \texttt{fluents(\#Fluents)}, where \#Fluents is a list of fluents. 

More details about the STAR system can be found in relevant work of Diakidou et al. \cite{Diakidoy2015} and on the system's website. The full example used in this section, is available at the ``Help" menu of the Web-STAR IDE. Readers interested in the technical aspects of the argumentation semantics of the STAR System are directed to work of Diakidou et al. \cite{Diakidoy2014}.

\section{A Web Interface For The STAR System (Web-STAR)}
\label{section:webstar}
The Web-STAR IDE is available online (http://cognition.ouc.ac.cy/star/) and is accessible from any device. Users are presented with a web interface that includes the source code editor, textual and visual output of the comprehension model, a number of control and debugging options and online help. 

Moreover, the IDE employs a number of social features for user collaboration, like public code sharing and posting of stories to the public repository. In addition, users can work together using a state-of-the-art collaboration utility that allows screen sharing, text and voice chat and presenter following functionality. In short, work on Web-STAR includes:
\begin{itemize}
\item A WEB-IDE that does not require setup, is OS agnostic and offers most modern IDEs' functionality
\item A platform for collaboration and teaching
\item A platform for integrating story comprehension functionality to other systems 
\end{itemize}

\begin{figure}[ht]
\begin{center}
\includegraphics[width=1.1\linewidth]{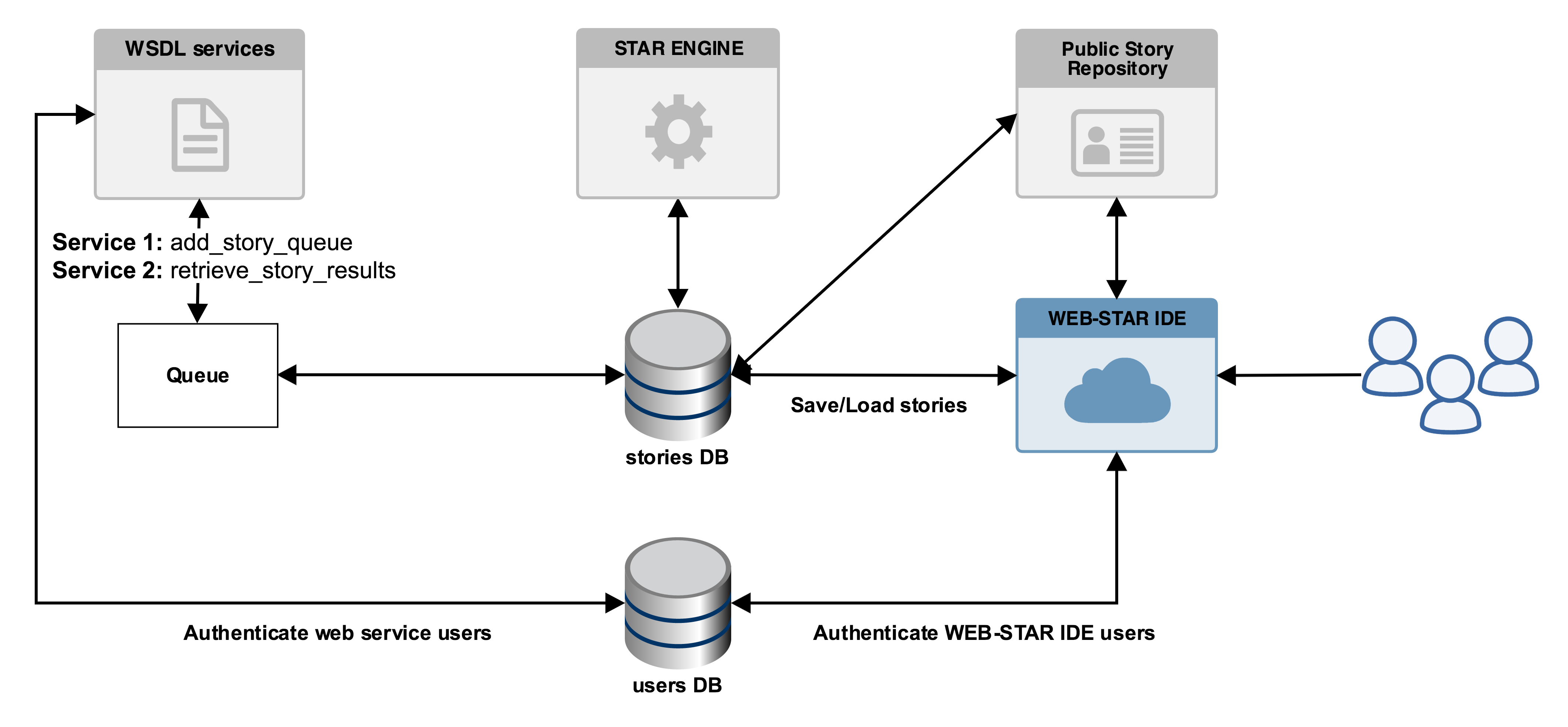}
\caption{The Web-STAR system architecture. The diagram includes the authentication mechanism, the storage functionality and the web services provided.}
\label{fig:webstar:architecture}
\end{center}
\end{figure}

Figure \ref{fig:webstar:architecture}, depicts the architectural diagram of the Web-STAR platform that comprises the Web-STAR IDE, the web services, the STAR system engine, the public repository and the databases for storing related information. In the next paragraphs, an example of using the web-based IDE is depicted using a scenario where ``Nick", a university student, wants to learn how to use the STAR system for running experiments on story comprehension.

\subsubsection{The IDE Environment and Workspace}
\label{subsection:workspace}
First, he creates an account to use the system and activates the personal workspace. Currently, both local and remote authentication options are available. The local authentication method uses the integrated storage facilities of the platform and remote authentication uses the OAuth2 protocol (https://oauth.net/2/), offered by third parties like Facebook, Google, Github, etc.

After the authentication process is complete, he is redirected to the source code editor, where he can write or load code and edit it. The source code editor is based on ACE editor (https://ace.c9.io), an open source web editor used also by many other popular cloud IDEs. This editor was chosen because of its maturity and open source license. Ace is a code editor written in JavaScript and includes features like syntax highlighting, theming, automatic indent and outdent of code, search and replace with regular expressions, tab editing, drag-drop functionality, line wrapping and code folding. Moreover, this editor can handle huge documents with more than one million lines of code.

\begin{figure}[ht]
\begin{center}
\includegraphics[width=1.0\linewidth]{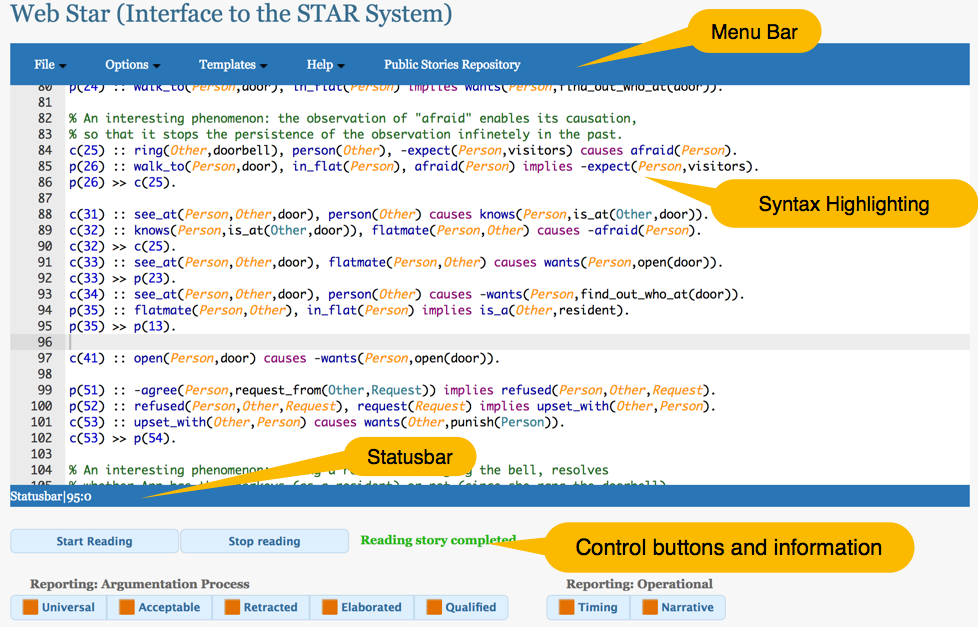}
\caption{A screenshot of the source code editor, depicting the line numbering, syntax highlighting and line highlighting. Above the source code editor resides the menu and at the bottom is the statusbar.}
\label{fig:webstar:code_editor}
\end{center}
\end{figure}

More specifically, users can write their code from scratch in the source code editor, load it from an external file or load an example file. Non-experts like Nick, can benefit from the many examples available. Currently, the source code editor allows syntax highlighting using a STAR syntax highlighter file that inherits the Prolog's syntax template. Furthermore, line numbering and code wrapping are also available to users. 

The Web-STAR IDE has a comprehensive list of menu options that enable users to load example story files, study them and edit them. Users have a personal workspace for saving their newly created programs and a public workspace for loading other user's programs. A program that is saved in the personal workspace can only be accessed by its creator, whereas a program stored in the public space, is visible to everyone. Options for importing code stored locally on the user's personal device and exporting stories to a file for local processing are also available. This functionality allows a user to use a locally installed version of the STAR system to process the story.

Nick, loads the example story presented in Section \ref{section:star_system}. The source code editor immediately identifies and highlights the STAR semantics (variables, rules, operators, etc.) and makes it easy for Nick to understand the example (see Figure \ref{fig:webstar:code_editor}). After studying the file, he moves to the questions part and he decides to try and add one more question by choosing the ``question template" from the menu. The question template (presented in the previous Section) is added and Nick adds the predicates and time-points where the question is posed. Since he has done changes to the example file, he now saves it to the personal workspace using the corresponding menu option.

\subsubsection{Story Comprehension and Output}
\label{subsection:story_comprehension}
Nick completes the changes in the story and he tries to process it. He clicks the ``start reading" button and he immediately sees results coming from the STAR system in the ``Raw output" tab. He chooses the Graphical output since its easier for a non-expert to follow the changes as the story progresses through the different time-points. 

Results include both the comprehension model and answers to the questions posed in the story file. Web-STAR presents its output both in textual and graphical format. The output area comprises two tabs, one for the raw output and one for the visual output. Each tab is dynamically updated when new information is sent from the STAR system.

\begin{figure}[ht]
\begin{center}
\includegraphics[width=1.0\textwidth]{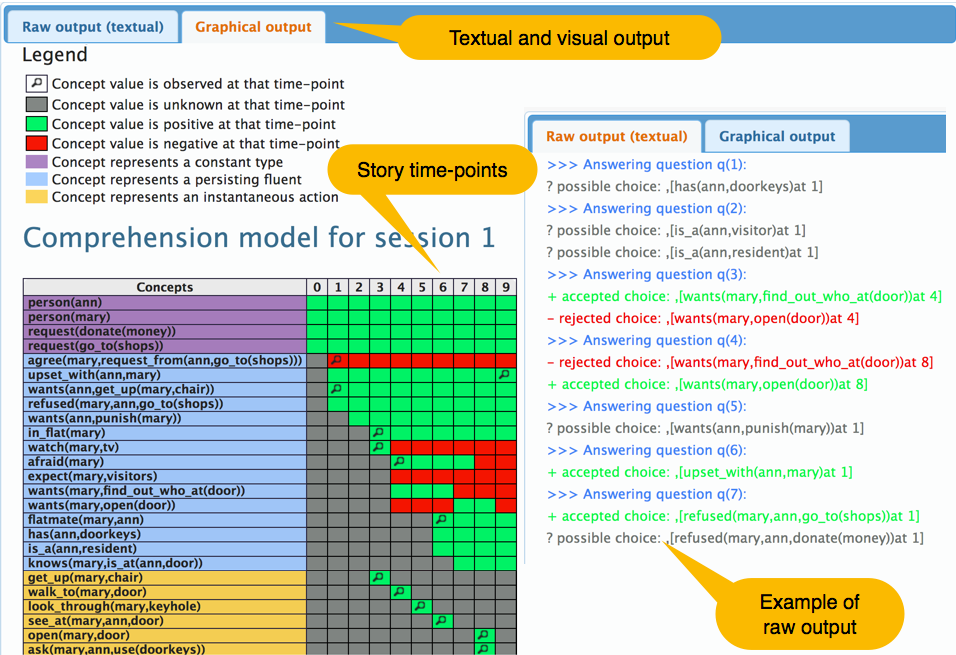}
\caption{A screenshot of the ``Graphical output" tab of the IDE, showing the comprehension model. The legend above the comprehension model provides details on the meaning of symbols and colors in the graphical model. On the right side, the raw output for the same story is presented.}
\label{fig:webstar:live_reasoning}
\end{center}
\end{figure}

In Figure \ref{fig:webstar:live_reasoning}, the graphical output of the comprehension model is depicted, presenting the state of each concept at each time-point. Green represents concepts whose value is positive at that time-point, red represents concepts whose value is negative at that time-point and dark grey represents concepts whose value is unknown at that time-point.
The magnifying glass is for concepts whose value is observed at that time-point, i.e., they are extracted from the narrative content and are not inferred. Concepts with orange background represent an instantaneous action, concepts with light blue background represent a persisting fluent and concepts with purple background represent a constant type (e.g., \texttt{person(ann) at always.}).

\subsubsection{Reporting Options}
\label{subsection:reporting_options}
Further studying on how the STAR system works for story comprehension, made Nick curious on the internal processes and more specifically on how the argumentation mechanism is applied for story comprehension. He decides to enable the reporting options. Users have the option to enable or disable reporting functionality used for debugging. In particular, users can choose to view all arguments (\texttt{Universal}), a subset of the acceptable arguments (\texttt{Acceptable}), arguments that are removed from a specific session (\texttt{Retracted}), arguments added in a specific session(\texttt{Elaborated}) and information about which arguments qualify other arguments (\texttt{Qualified}).

\subsubsection{Collaboration and ``Social" Options}
\label{subsection:collaboaration}
\begin{figure}[ht]
\begin{center}
\includegraphics[width=1.0\textwidth]{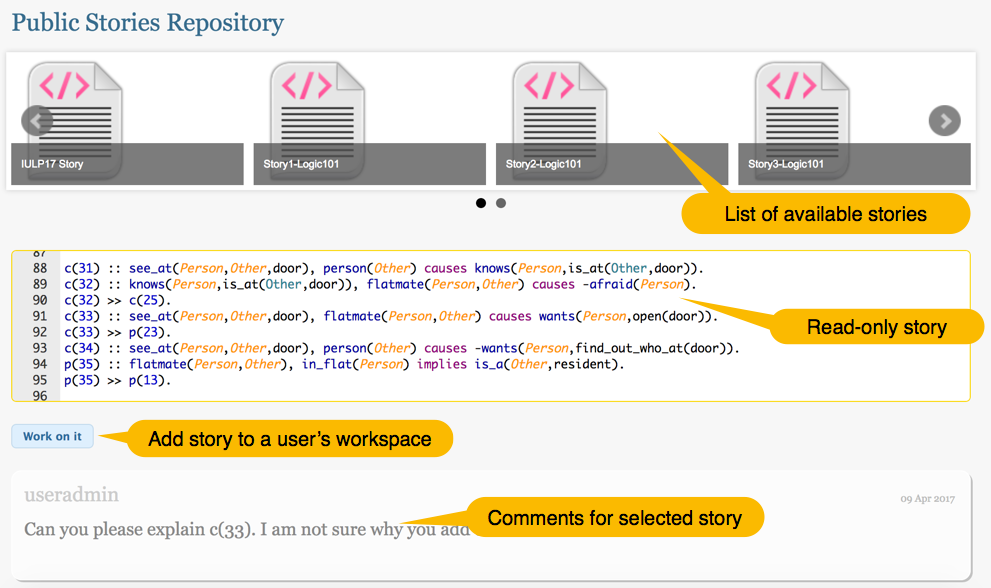}
\caption{A screenshot of the Public Stories Repository. Users can add comments on stories and ask questions. If interested, they can start working on a story by copying it to their personal workspace.}
\label{fig:webstar:public_repository}
\end{center}
\end{figure}

Apart form the ``expected" IDE functionality, Web-STAR also provides functionality for sharing publicly a story file for others to use as an example. Users can share a story by clicking the ``Share it" button. This story becomes visible to the public stories repository (see Figure \ref{fig:webstar:public_repository}) and the public stories menu. Users can add comments on a story and read it using the built-in functionality. This is used primarily for educational purposes, from users that are trying to learn and would benefit from the opinions of other more expert users.

Moreover, users can collaboratively write a story using the collaboration functionality provided. The system produces a link that can be send to anyone interested in collaborating for a specific session through an email or an instant messaging application. Anyone with the link can see the screen, mouse pointers of each participating user, changes of content in real-time and can chat using both text and audio. This setting enables teams to collaborate for creating a story file and students for learning and working together on class projects.

Nick thinks that the story he created can be useful to other students and he shares it publicly by clicking on the ``Share it" button. Not long after, some of his classmates added some comments and a discussion around the story starts.

\subsection{System Implementation}
\label{subsection:system_implementation}
The system is based on PHP for backend operations, MariaDB database for data storage and JQuery javascript library for the front-end design. Behind this infrastructure lies the STAR system \cite{Diakidoy2015} and the Prolog interpreter. A wrapper is employed for sending the story file from the front-end to the back-end and returning the results in real-time from the Prolog interpreter using the HTML5 ``Server-Sent Events" functionality to dynamically update the interface.

All data storage is handled through the MariaDB database. In particular, a number of tables are used for storing user data, user profiles and the STAR web service queue.

For the front-end design, HTML, JavaScript and CSS3 are used to present an easy to use web environment. The JQuery library (https://jquery.com/) is also used to add intuitive UI components and AJAX functionality. Collaboration functionality is provided using both AJAX components for sharing and commenting on stories and the TogetherJs library (https://togetherjs.com/).

\subsubsection{Web Services}
\label{subsection:web_services}
The Web-STAR platform exposes two web services that can be used by third party applications for adding a domain file to the STAR system queue for processing and retrieving the results of the reasoning process when completed (see Figure \ref{fig:webstar:architecture}). This approach was chosen to minimize the waiting time in cases of large story files that require extensive processing.

\subsection{Platform Usage Metrics and Examples}
\label{subsection:current_usage}
The Web-STAR platform is available online. Currently, more than thirty users have registered and contributed more than 50 stories. Moreover, the system has received more than 1350 web service calls for processing STAR domain files.

The platform was also used in the Robot Trainer Game \cite{Rodosthenous2016} for real time processing of the acquired knowledge rules. Robot Trainer is a Game With A Purpose (GWAP) that uses a hybrid approach for acquiring knowledge in the form of rules using both human contributors and the Web-STAR platform. Moreover, these rules were evaluated by humans for their applicability on answering questions about stories different than the ones used to generate them.

\section{Ongoing Work}
\label{section:ongoing_work}
Moving towards the direction of maximizing the number of people that are able to use a story comprehension system, we are currently working on a number of visualization additions to the Web-STAR IDE, which could eventually be used in other domains as well. Some of these additions are already developed to match the needs of other systems and are currently being integrated to the Web-STAR IDE. In the following paragraphs, these additions are presented with the relevant discussion on the advantages of using them. Currently, we are still investigating the various approaches that could be used in this direction.

\subsection{Visualizing the Background Knowledge}
\label{subsection:visualizing_bk}
\begin{figure}[!ht]
\begin{center}
\includegraphics[width=0.8\textwidth]{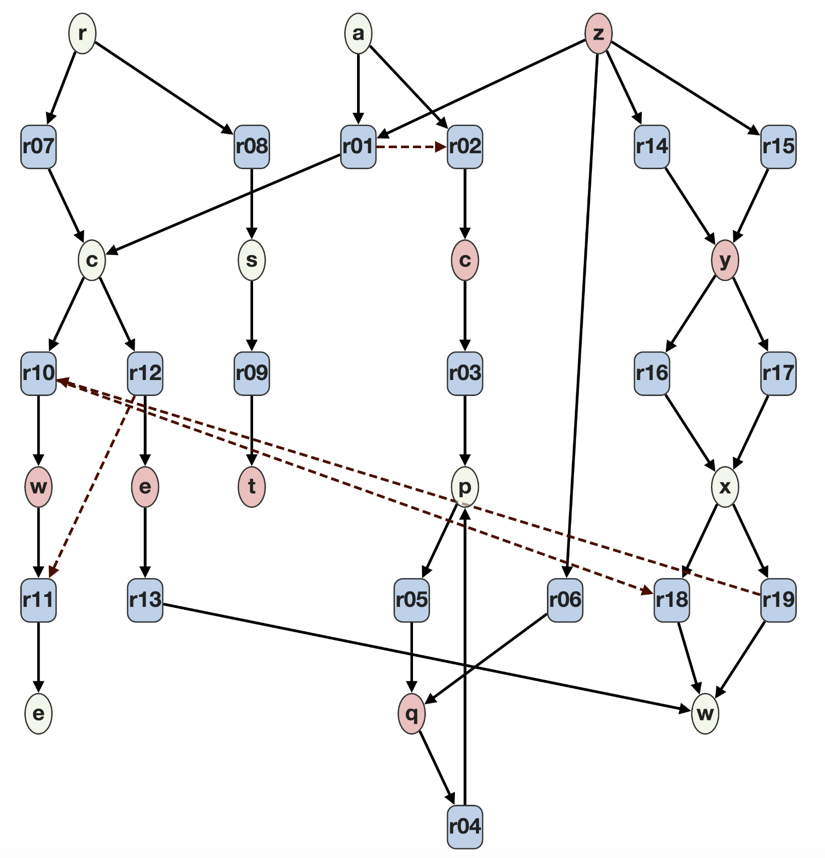}
\caption{An indicative example of a directed graph that can be used for representing visually the background knowledge. This screenshot is from a prototype version of the Web-STAR IDE that is under development and will soon be released. More specifically, the positive and negative concepts are presented using green and red nodes respectively and preferences between rules are presented with a dashed edge between nodes.}
\label{fig:future_visualizations:rules}
\end{center}
\end{figure}

Some of the processes that could be visualized include the previewing, addition and editing of rules. For these processes a directed graph with nodes and edges could be used. For example, the rules presented in the knowledge base snippet below can be represented by the directed graph in Figure \ref{fig:future_visualizations:rules}:
\begin{quote}
\texttt{r01 :: a, -z implies c.}\\
\texttt{r02 :: a implies -c.}\\
\texttt{r03 :: -c implies p.\\}
\texttt{$\dots$\\}
\texttt{r01 >> r02.\\}
\texttt{r19 >> r10.}
\end{quote}

Users are able to change a rule by just clicking on the corresponding edge or node in order to remove or edit it. Users also have the ability to observe rules that build on each other i.e., rules that the predicates in the head exist in another rule's body, and better understand the reasoning process. For adding a rule, users will still have to type a compatible with the STAR system semantics name and add predicates for the head and body of the rule.

Since some of these story files might grow big, it would be useful for the user to isolate the part of the background knowledge that wants to work with at a given time. To support this, a type of code folding functionality needs to be implemented, which will allow users to focus on a specific group of rules in the screen and ``minimize" the others, either by hiding the head and body of each rule or by showing them in a smaller size.

This functionality can also be used for feedback on the reasoning process. For example, a user can choose to change the priority between two rules by just clicking on the edge that connects them to remove it and add a new edge between different rules. Such functionality could be useful in other domains, where users need to give their feedback and change the knowledge rules and priorities stored in the system.

In particular, we are examining the addition of this functionality in a Game platform, where users will train their ``Player" on what strategy to choose for winning the game. This platform will enable game players to visually add rules and priorities without any prior knowledge of the underlying reasoning system.

\subsection{Support for Natural Language Input}
\label{subsection:towards_an_automated_ide}
Aiming in a fully automated solution, a number of problems must be addressed before deploying such a system. These problems involve the encoding of natural language text to symbolic form suitable for the STAR system, event extraction from text, temporal event ordering and acquisition of background knowledge suitable for a selected story.

A number of these problems, can be addressed by incorporating well established systems to the IDE's toolset. Systems like PIKE \cite{2016tkde} that is able to extract knowledge from natural language into an RDF triplet using Semantic Role Labeling \cite{UzZaman2013} and Stanford CoreNLP \cite{manning-EtAl:2014:P14-5} that can process a text in natural language and extract knowledge information, are good candidates for this task. Furthermore, there are a number of techniques that can be used for temporal event ordering \cite{UzZaman2013}. More difficult tasks, like background knowledge acquisition, can be sourced to the crowd, using Games With A purpose \cite{VonAhn2008} or other crowd based techniques, or can be addressed through machine learning approaches \cite{Michael_2009_ReadingBetweenTheLines}. Even Knowledge Bases like Conceptnet \cite{Speer2013}, YAGO \cite{Mahdisoltani2015} and NELL \cite{NELL-aaai15} can be used for this task or even manually coded rules from the OpenCyc project \cite{Lenat1995}.

\section{Conclusions}
\label{section:conclusions}
In this work, we described the Web-STAR platform, built on top of the STAR system for story comprehension. Information on the technologies used for implementing the platform, the user interface and the collaboration tools available are presented. Moreover, a number of examples were used to explain how to use the Web-STAR IDE to prepare a story and a use case scenario was used to elaborate on how to process a story. The platform is currently used by University students and Instructors for educational purposes and as part of a GWAP, which benefits from the platform's web services for answering questions.

Ongoing work on this platform aims in enabling non-expert users to easily use and experiment with the STAR system. These additions will also provide the basis of the feedback module needed for integrating the STAR or other similar engine with other systems.
\newpage
\bibliographystyle{splncs03}
\bibliography{iulp-1}

\end{document}